\documentclass[12pt]{article}

\usepackage[utf8]{inputenc}
\usepackage{times}
\usepackage{subcaption}
\usepackage{booktabs}

\usepackage{graphicx}% Include figure files
\usepackage{dcolumn}% Align table columns on decimal point
\usepackage{bm}% bold math
\DeclareFontFamily{OT1}{mathc}{}
\DeclareFontShape{OT1}{mathc}{m}{it}{<-> mathc10}{}
\DeclareMathAlphabet{\mathabxcal}{OT1}{mathc}{m}{it}
\usepackage[superscript]{cite}
\usepackage{lineno}
\usepackage{physics}
\usepackage{newtxmath}
\usepackage{upgreek}
\usepackage{gensymb}
\usepackage{units}
\usepackage{float}
\usepackage{multirow}
\usepackage{xr}
\usepackage[inkscapelatex=false]{svg}
\usepackage[colorlinks=true,linkcolor=blue,urlcolor=blue,citecolor=blue]{hyperref}
\usepackage[commentmarkup=uwave,deletedmarkup=sout]{changes}
\definechangesauthor[name=Malte Grosche, color=blue]{fmg}
\makeatletter
\renewcommand\@biblabel[1]{#1.}
\makeatother

\usepackage{gensymb}
\usepackage{mathtools}
\usepackage{caption}
\newenvironment{sciabstract}{%
\begin{quote} \bf}
{\end{quote}}

\title{\vspace{-1cm} 3D bulk-resolved $g$-wave altermagnetic order parameter in CrSb}

\author{Mengmeng Long,$^{1\dagger}$ Theodore I. Weinberger,$^{1\dagger\#}$ Zheyu Wu,$^{1\dagger}$\\ Mads F. Hansen,$^{1}$ Ran Tao,$^{1}$ Mridul Shrestha,$^{1}$ Dave Graf,$^{2}$\\ Yurii Skourski,$^{3}$ F. Malte Grosche,$^{1}$  Alexander G. Eaton$^{1\ast}$
\\
\normalsize{$^{1}$Cavendish Laboratory, University of Cambridge,}\\
\normalsize{JJ Thomson Avenue, Cambridge, CB3 0US, UK}\\
\normalsize{$^{2}$National High Magnetic Field Laboratory, Tallahassee, Florida 32310, USA}\\
\normalsize{$^{3}$Hochfeld-Magnetlabor Dresden (HLD-EMFL), Helmholtz-Zentrum }\\ \normalsize{Dresden-Rossendorf, Dresden, 01328, Germany}\\
\\
\normalsize{$^\dagger$These authors contributed equally to this work.}
\\
\normalsize{$^\#$tiw21@cam.ac.uk}\\
\normalsize{$^\ast$alex.eaton@phy.cam.ac.uk}
}

\date{\today}

\begin{document}

\baselineskip24pt

\maketitle

\clearpage
%\linenumbers
\begin{sciabstract}

Electronic phases of matter, such as magnetism and superconductivity, are defined and distinguished by their order parameters quantifying the spontaneous symmetry breaking underlying each phase. Simple cases include the uniform magnetisation of ferromagnets~\cite{curie1895proprietes,heisenberg1928} and isotropic gap function of conventional superconductors~\cite{BCS}. Unconventional superconductors~\cite{Stewart_review_doi:10.1080/00018732.2017.1331615} often have a nodal gap function, where the gap changes sign at nodes on the Fermi surface. This concept of unconventional/nodal order parameter symmetry has recently been extended to numerous magnetic systems~\cite{jungwirth2024altermagnetsbeyondnodalmagneticallyordered,liu2025different,song2025electrical,yamada2025metallic}, including altermagnets~\cite{BeyondPRX.12.031042,EMERGINGPRX.12.040501,hayami2019momentum,smejkal2020sciadv,ma2021multifunctional}, in which up- and down-spin species are non-degenerate around the Fermi surface. Here we demonstrate that magnetic quantum oscillation~\cite{Shoenberg1984} measurements can provide a high resolution, bulk-sensitive, 3D mapping of the order parameter in an unconventional magnet. By rotating a magnetic field through high- and low-symmetry directions of the CrSb Brillouin zone, we show that this material's altermagnetic band structure leads to a reduction of symmetry for each spin-split Fermi sheet away from nodal orientations. In momentum space, the exchange splitting between up and down spins follows the profile of the $\mathcal{Y}_{4}^{-3}=zy(3x^2-y^2)$ real spherical harmonic -- analogous to a $g$-orbital of the hydrogen atom. While notoriously difficult to resolve in unconventional superconductors, our work demonstrates that the order parameter symmetry of unconventional magnets can be precisely mapped via quantum-oscillatory quasiparticle spectroscopy, establishing CrSb as a prototypical $g$-wave metallic altermagnet.

\end{sciabstract}

\clearpage

\noindent
There are two traditional types of collinear magnetic ordering: ferromagnetism~\cite{curie1895proprietes,heisenberg1928}, in which all spins point in the same direction, and antiferromagnetism~\cite{neel1932}, in which the spin polarisation alternates up and down from site to site. In an antiferromagnet, the magnetic structure can be decomposed into two magnetic sublattices, resulting in a doubling of the magnetic unit cell~\cite{ShullPhysRev.76.1256.2} under translation. Recently, a third distinct classification was proposed, dubbed \textit{altermagnetism}~\cite{BeyondPRX.12.031042,EMERGINGPRX.12.040501,hayami2019momentum,smejkal2020sciadv,ma2021multifunctional}. Like antiferromagnets, altermagnets possess alternating up and down spin orientations from site to site. However, the distinction is that in an altermagnet the spin sublattices are connected by \textit{rotational} symmetries rather than just translation (or inversion). This leads to the lifting of Kramers spin degeneracy~\cite{kramers1930} and a momentum-dependent non-relativistic spin-splitting of the electronic band structures~\cite{jungwirth2025altermagnetism}. Therefore, altermagnetism represents a distinct form of ordering, in which the momentum-dependent spin-split property of ferromagnets is combined with the spin-compensated zero net magnetisation of antiferromagnets, opening new opportunities for magnetic memory and spintronic device applications~\cite{song2025altermag-rev}.

Numerous altermagnet candidates have been identified from \textit{ab initio} calculations,~\cite{BeyondPRX.12.031042,EMERGINGPRX.12.040501,fender2025altermagnetism,song2025altermag-rev,WanPRL25} with experiments on MnTe giving good empirical correspondence with theoretical expectations~\cite{krempasky2024altermagnetic,Lee-MnTePRL24,Osumi-MnTePRB24,amin2024nanoscale}. However, while many semiconducting or insulating materials have been identified, only a handful of metallic altermagnet candidates have so far been proposed. The realisation of a metallic altermagnet, with an ordering temperature far above 300~K, would be particularly desirable for efficient electronic transport in technological device settings~\cite{WanPRL25}.

Of metallic candidates, surface-sensitive photoemission spectra of RuO$_2$~\cite{fedchenko2024TRSBSciAdv,lin2024observationgiantspinsplitting}, KV$_2$Se$_2$O~\cite{jiang2025metallic} and CrSb~\cite{reimers2024direct,DingPhysRevLett.133.206401,zeng2024observation,yang2025three} have been interpreted to show hallmarks of altermagnetic spin-splitting. However, for the case of RuO$_2$ a variety of subsequent bulk-sensitive measurements~\cite{HiraishiPRL24,kessler2024absence,Tony2025RuO2} cast considerable doubt on whether this material is indeed an altermagnet. Further photoemission studies~\cite{Liu-PRL.133.176401,Osumi-RuO2PRB26} showed that this material appears to possess a topological surface state with Rashba-like spin splitting, which may give the appearance of altermagnetically-lifted Kramers spin degeneracy even though the bulk is actually nonmagnetic. A similar story has unfolded with KV$_2$Se$_2$O, for which the surface termination layer may indeed be altermagnetic~\cite{lange2026emergentaltermagnetismsurfacesantiferromagnets}, yet neutron diffraction clearly resolves conventional antiferromagnetism throughout the bulk~\cite{Sun_KV2Se2OAFM-PRB25}. These cautionary tales motivate bulk-sensitive studies of candidate altermagnets, in order to firmly establish their intrinsic magnetic properties.

\subsection*{Nodal magnetic order parameters}

We can frame our thinking of magnetic order parameter symmetries by analogy to unconventional superconductors. In a conventional $s$-wave BCS superconductor, the order parameter is the gap function $\Delta(\vec k)$, which is an isotropic energy separation corresponding to a dispersion possessing the same symmetries as the crystal. By contrast, unconventional superconductors~\cite{Stewart_review_doi:10.1080/00018732.2017.1331615} are those in which the pairing symmetry breaks crystal point group symmetries, or where the gap function exhibits a non-trivial phase structure (like a sign change) not required by the crystalline symmetry.

These concepts have recently been applied to magnetic systems~\cite{jungwirth2024altermagnetsbeyondnodalmagneticallyordered,liu2025different}. Let us define the order parameter for an unconventional magnet as the exchange splitting (energy difference) between up- and down-spin states at a given wavevector, $\Delta(\vec k) = E(\vec k, \uparrow) - E(\vec k, \downarrow)$. This produces spin-split Fermi sheets separated by the (perpendicular) wavevector $k_\perp \simeq \Delta(\vec k)/(\hbar v_{\text{F}})$, where $v_{\text{F}} = \partial E/(\hbar \partial k) $ is the Fermi velocity. In this way, a ferromagnet is analogous to a BCS superconductor -- its nonzero magnetisation splits the Fermi surface into unequal majority and minority spin species, characterised by an isotropic energy gap between them. It follows that for an unconventional magnet of $p$/$d$/$f$/$g$-wave symmetry, the Fermi surface will possess 1/2/3/4 highly symmetric nodal planes at which up-character swaps to down, and vice versa.

%In an altermagnet, the volume enclosed by the Fermi pocket of one spin species must be equal to that of its opposite spin counterpart, and the overall order parameter symmetry is even in parity (i.e. $d$-, $g$- or $i$-wave).

%the difference (in momentum space) between up and down spin-split Fermi sheets, i.e. $\Delta(\vec k) = \vec k_{\text{F}}^\text{up} - \vec k_{\text{F}}^\text{dn}$ where $k_{\text{F}}$ is the Fermi wavevector. In this way, a ferromagnet is therefore directly analogous to a BCS superconductor -- its nonzero magnetisation splits the Fermi surface into unequal portions of majority and minority spin species, with an isotropic separation in $\vec k$ between them. It follows that for an unconventional magnet of $p$/$d$/$f$/$g$-wave symmetry, the Fermi surface will possess 1/2/3/4 highly symmetric nodal planes at which up-character swaps to down, and vice versa. In an altermagnet, the volume enclosed by the Fermi pocket of one spin species must be equal to that of its opposite spin counterpart, and the overall order parameter symmetry is even in parity (i.e. $d$-, $g$- or $i$-wave).

\begin{figure}[!htbp]
\vspace{-2.5cm}
\begin{center}
\includegraphics[width=1\linewidth]{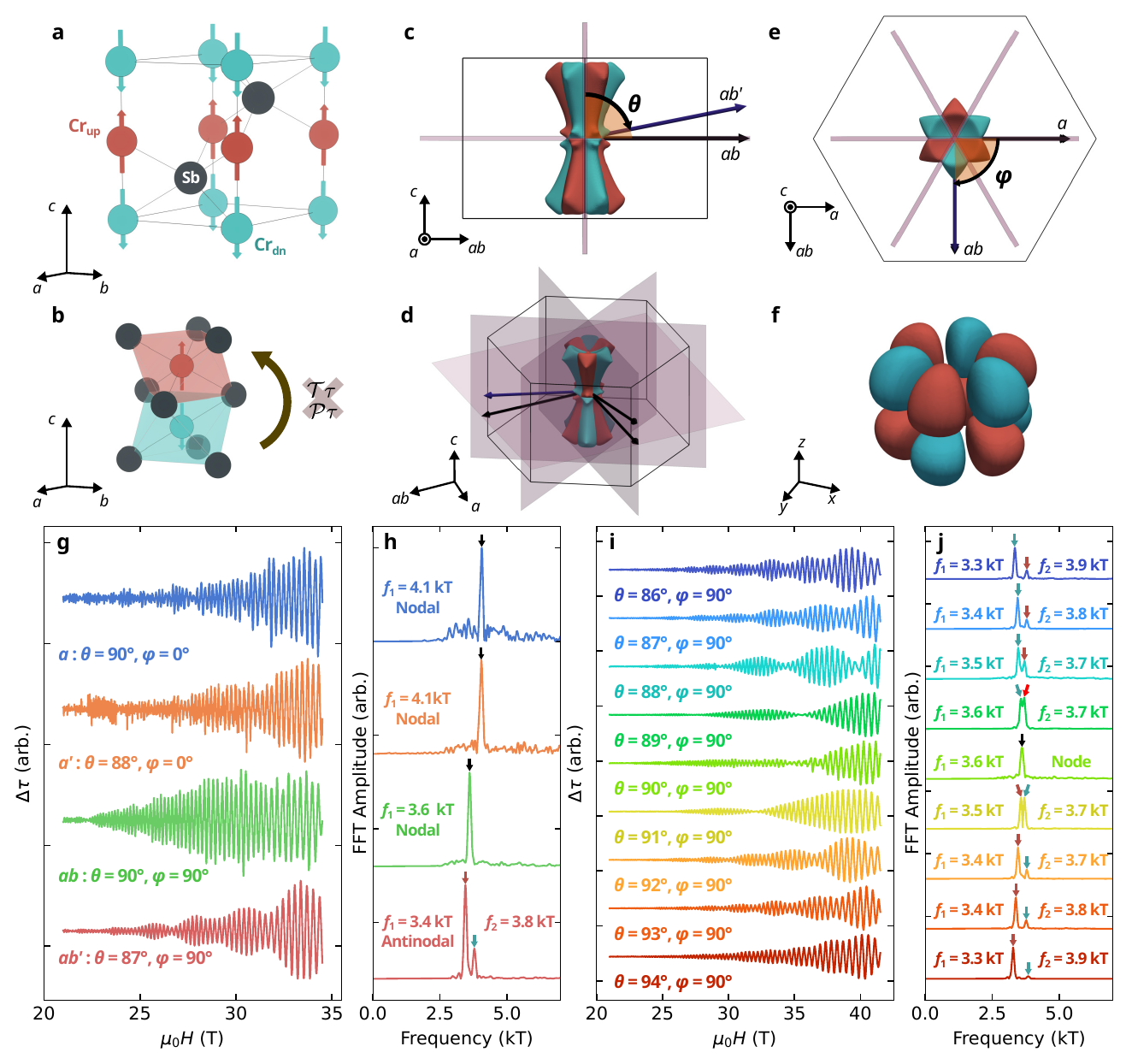}
\end{center}
\caption{\textbf{Nodal planes bisect $g$-wave-symmetric spin-split Fermi surface sheets in CrSb. a,} The crystal structure of CrSb, with alternating magnetic moments on the Cr sites oriented along the $c$-axis. Red (blue) indicates spin-up (spin-down). \textbf{b,} The trigonal arrangement of the antimony ions means that mapping a red chromium site to a blue one necessarily requires a screw rotation. \textbf{c-e,} The primary Fermi surface sheet of CrSb is a closed 3D ellipsoidal pocket, shaped like a dogbone. Nodal planes -- where spin degeneracy is imposed by symmetry -- are given by thin shaded slabs. The azimuthal angle $\varphi$ is defined as the inclination from $a$ to $ab$ in the hexagonal basal plane; the polar angle $\theta$ is from $c$ towards the basal $ab$ plane. \textbf{f,} Visual depiction of the $\mathcal{Y}_{4}^{-3}=zy(3x^2-y^2)$ real spherical harmonic, which characterises the $g$-wave symmetry profile of altermagnetic spin-splitting in CrSb.\parfillskip=0pt}
\label{Fig1}
\end{figure}
\nopagebreak
\begin{figure}[!htbp]
    \ContinuedFloat
    \begin{centering}
    \caption{(cont.) \textbf{g,} Quantum oscillations in the background-subtracted magnetic torque $\Delta \tau$, rescaled to the same maximal amplitude and high-pass filtered for ease of presentation (see \textit{Methods} for details) at magnetic field \textbf{H} orientations as indicated, and \textbf{h,} their fast Fourier transform (FFT) frequency spectra. A singular FFT peak is observed in the three nodal orientations, which splits into two peaks in the antinodal plane. All measurements were performed at 0.4~K. \textbf{i,j} $\Delta \tau$ and corresponding FFT spectra for small rotations of $\theta$ in the antinodal plane away from the $ab$ direction towards $c$. For \textbf{H} aligned along $ab$, only one frequency is observed, which elsewhere splits into two distinct peaks, indicating altermagnetically spin-split Fermi sheets.}
    \end{centering}
\end{figure}

CrSb crystallises in the hexagonal P6$_3$/$mmc$ NiAs-type structure, with chromium atoms stacked along the crystallographic $c$-axis in octahedrally coordinated layers, while the antimony atoms fill the interstitial sites in a trigonal prismatic arrangement~\cite{willis1953crystal} (Fig.~\ref{Fig1}a,b). Below an ordering temperature of around 740~K (see Extended Data Fig.~1), the chromium sites possess alternating magnetic moments oriented along the $c$-axis~\cite{Snow52PhysRev.85.365}. The trigonal arrangement of antimony necessarily requires a $6_3$ screw rotation to map a spin-up chromium site to its spin-down counterpart. Consequently, while CrSb breaks both pure time-reversal and primitive lattice-translation symmetries, it preserves a \textit{combined symmetry} pairing 
this non-symmorphic spatial operation with time reversal. These symmetry properties motivate the proposal~\cite{EMERGINGPRX.12.040501,WanPRL25,ma2021multifunctional,yang2025three,Prag2025CrSb-NatComm,singh2025chiralspinsplitmagnonsmetallic} that CrSb is a metallic altermagnet in which $\Delta(G\vec k) = -\Delta(\vec k)$, where $G$ is this composite operation.

\subsection*{Quantum oscillation measurements}

In this work, we map the symmetry of $\Delta(\vec k)$ in CrSb by performing magnetic quantum oscillation (QO) measurements through the de Haas-van Alphen effect~\cite{Shoenberg1984,dHvA}, and show that this material is a $g$-wave altermagnet. QO experiments are an especially well-suited technique for resolving the nodal planes of an unconventional magnet, and hence the order parameter symmetry. This is because $\Delta(\vec k)$ does not just go to zero for $\vec k$ on a nodal plane -- it is also mirror-symmetric about the plane. Aligning a magnetic field within a nodal plane therefore produces QO orbits for up- and down-spin Fermi sheets that are different, but can be mapped onto each other by a mirror operation about the nodal plane, and which therefore enclose the same area. This leads to a single QO frequency per Fermi pocket for fields in nodal planes. By contrast, for fields oriented away from nodal planes, the up- and down-spin Fermi sheets are not related by any symmetry operation, and therefore typically enclose \textit{different} areas, leading to two distinct QO frequencies. This thereby yields an expected smoking-gun signature of altermagnetic spin-splitting.

Throughout this article, we shall refer to rotations of the orientation of an applied magnetic field \textbf{H} through azimuthal angles $\varphi$ defined in the crystallographic $ab$ plane, and polar angles $\theta$ between $c$ and the $ab$ plane (Fig.~\ref{Fig1}c,e). Note that for hexagonal CrSb the $a$ direction is equivalent to $b$. The $c-a$ plane is a highly symmetric nodal plane in which Kramers degeneracy is enforced by symmetry, while $c-ab$ is antinodal, in which spin degeneracy is only symmetry-enforced at $c$ and $ab$. In CrSb, $g$-wave splitting should result in nodal planes every $\delta\varphi = 60\degree$ for rotations in the $a-ab$ plane, and every $\delta\theta=90\degree$ rotating through the $c-ab$ plane.

We present the experimentally deduced central Fermi surface pocket of CrSb, with its high-symmetry nodal planes between up and down sheets, in Fig.~\ref{Fig1}c-f. This sheet is a closed ellipsoid-like pocket of hole character, shaped like a dogbone. In Fig.~\ref{Fig1}g we plot background-subtracted magnetic torque $\Delta \tau$ as a function of $H$, to focus on the oscillatory component, measured by cantilever beam magnetometry (see \textit{Methods}). The blue waveform was measured for \textbf{H}  $\parallel a$ at $\theta = 90\degree, \varphi = 0\degree$, with the orange data collected a small rotation away in the $c-a$ plane to $\theta = 88\degree, \varphi = 0\degree$. The corresponding fast Fourier transform (FFT) spectra in panel \textbf{h} for both of these orientations exhibit a singular frequency peak at 4.1~kT.

\subsection*{Tracking $\Delta(\vec k) = E(\vec k, \uparrow) - E(\vec k, \downarrow)$}

Let us now contrast this with the case of fields in the $c-ab$ (antinodal) plane. The green curve of Fig.~\ref{Fig1}g was obtained for \textbf{H} $\parallel ab$ (i.e. $\theta = 90\degree, \varphi = 90\degree$), with a monofrequency profile of 3.6~kT. Upon a small $\theta$ rotation of 3$\degree$ towards $c$, a distinct beat pattern emerges (red curve), with two peaks clearly resolved in the frequency spectra. This is characteristic of the Fermi surface becoming spin-split into two non-degenerate sheets -- one spin-up and the other spin-down.

Fig.~\ref{Fig1}i,j tracks this split peak structure for fields close to the $ab$ direction. Incrementing $\theta$ in 1$\degree$ steps through the $c-ab$ plane, we find that the singular on-axis peak of 3.6~kT clearly splits into two distinct frequency branches, which by $4\degree$ of rotation are separated by 0.6~kT. This is reflected in the profile of the waveforms themselves, which exhibit a pronounced beat structure that is acutely sensitive to $\theta$.

\begin{figure}[!htbp]
\vspace{-0cm}
\begin{center}
\includegraphics[width=1\linewidth]{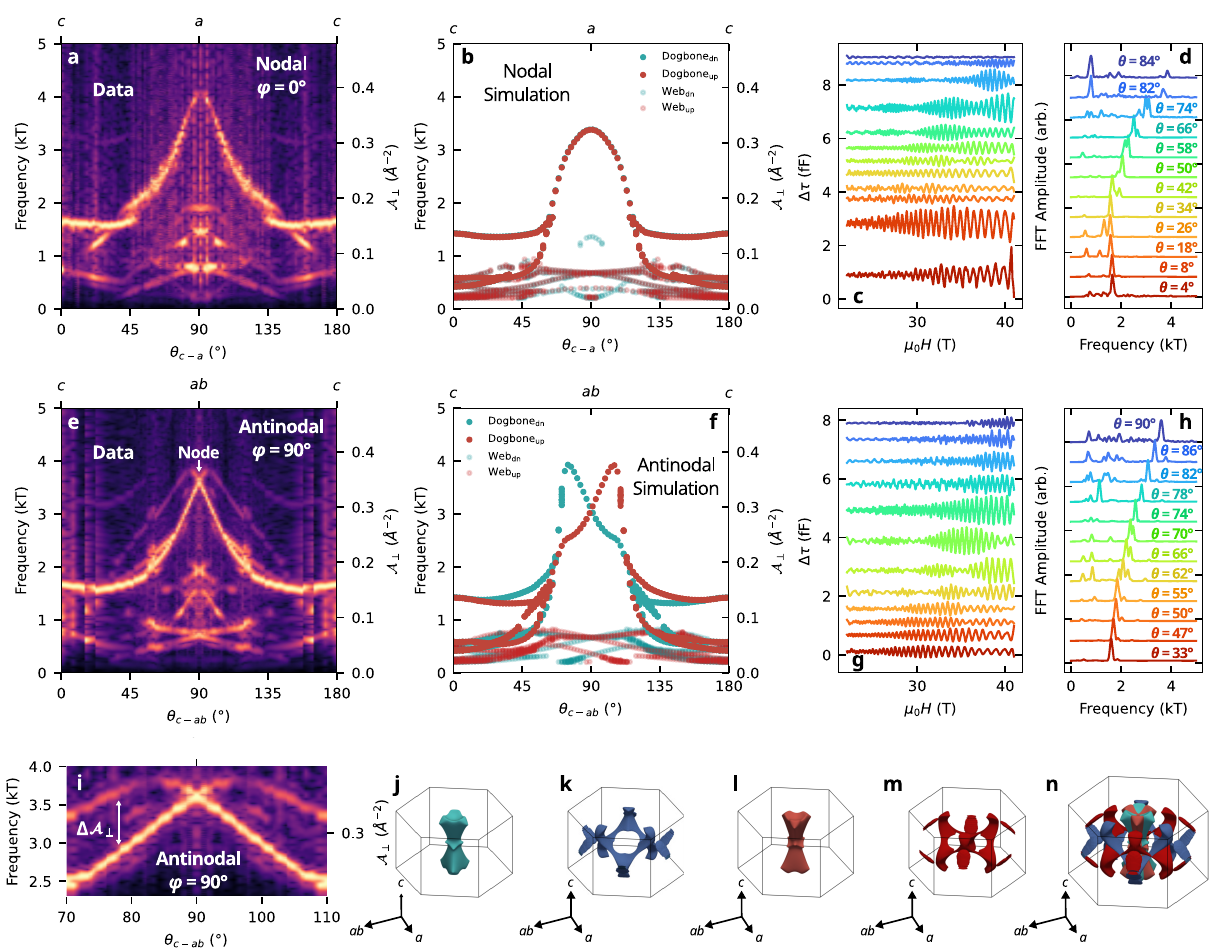}
\end{center}
\caption{\textbf{Mapping the altermagnetically spin-split Fermi surface of CrSb. a,} Heatmap of the spectral intensity of quantum oscillatory frequencies versus rotation angle in the nodal $c-a$ plane. Brighter heatmap intensities correspond to larger quantum oscillatory amplitudes; data have been symmetrised through 90$\degree$ for ease of presentation. \textbf{b,} Simulated frequency profile for the $c-a$ plane from DFT calculations (see \textit{Methods} and \textit{Supplementary Information}). \textbf{c,d} Selected $\Delta \tau$ curves and normalised FFTs that contributed to the heatmap in \textbf{a} (full dataset in Extended Data Fig. 2). All data were collected at 0.4~K. \textbf{e,} Data and \textbf{f,} Simulation for the antinodal $c-ab$ rotation plane, with waveforms and normalised FFTs in \textbf{g,h} (full dataset in Extended Data Fig. 3). Here spin-splitting between non-degenerate Fermi sheets is predicted by the simulation, which is clearly resolved experimentally close to $\theta = 90\degree$. In \textbf{i,} we present a zoomed-in view having high-pass filtered the raw data, to clearly show the non-degenerate spin-up and -down frequency branches crossing the nodal plane at $\theta = 90\degree$, which are separated by $\sim$1~kT. \textbf{j,k} Fermi surface rendering of the spin-down sheets of the dogbone and web pockets and \textbf{l,m} their spin-up counterparts. \textbf{n,} The full Fermi surface of CrSb, including all spin-up and spin-down sheets.}
\label{Fig2}
\end{figure}

The full angular profiles of QO frequency spectra for the $c-a$ (nodal) and $c-ab$ (antinodal) planes are shown in Fig.~\ref{Fig2}. QO frequencies are directly proportional to the Fermi surface cross-sectional area $\mathcal{A_\perp}$ normal to \textbf{H}~\cite{onsager_rel,LK}. We compare with simulated QO frequency spectra from density functional theory (DFT) calculations (see \textit{Methods} for details). The dogbone pocket exhibits QO frequencies of 4.1~kT for \textbf{H} $\parallel a$ and 3.6~kT for \textbf{H} $\parallel ab$. Two other bands of electron character also cross the Fermi level, leading to another more geometrically complex Fermi sheet with a web-like shape (see Fig.~\ref{Fig2}). The web sheet produces a very complicated dependence of QO frequency on angle in the low frequency range $\lessapprox$~2~kT. Away from nodal planes, the web is also split into up and down sheets, but its geometrical complexity makes this more difficult to resolve than for the dogbone. We therefore predominantly focus on the dogbone for \textbf{H} oriented close to $a$ or to $ab$, which produces QO orbits with frequencies $\gtrapprox$~3~kT, well above those from the web.

%hpf 2500 T in last fig, loess window 1.2 and poly 2,

Symmetry-enforced spin degeneracy causes the QO frequencies from both the spin-up and spin-down Fermi sheets to be equivalent for fields in the nodal $c-a$ plane, as depicted by our simulation (Fig.~\ref{Fig2}b). By contrast, for fields in the antinodal $c-ab$ plane, a large frequency splitting should occur over the approximate angular interval $60\degree < \theta < 120\degree$ (Fig.~\ref{Fig2}e,f), before the geometric profile of the dogbone leads to an accidental degeneracy. This is exactly what we observed in Fig.~\ref{Fig1}i,j and is clearly visible in the spectral intensity near $\theta = 90\degree$ in Fig.~\ref{Fig2}e. In Fig.~\ref{Fig2}i we present a zoom-in of this region, to focus on this spectral range. The overlap of frequencies at \textbf{H} $\parallel ab$, which is abruptly split upon incrementing $\theta$, is characteristic of the altermagnetic lifting of spin degeneracy, producing two spin-split Fermi sheets. Notably, whereas in the nodal plane both spin-up and spin-down frequency vs angle traces are expected to repeat every 90$\degree$, in the antinodal plane this reduces to every 180$\degree$ -- exactly as observed experimentally. 
%This unusual geometric dependence of frequency upon angle -- with a steep initial increase of one branch, with a maximum around $\theta = 96\degree$ before decreasing under further rotation -- is in excellent agreement with the DFT prediction of altermagnet spin-splitting of the band structure, which is most pronounced in this antinodal plane.

\begin{figure}[!htbp]
\vspace{-0cm}
\begin{center}
\includegraphics[width=1\linewidth]{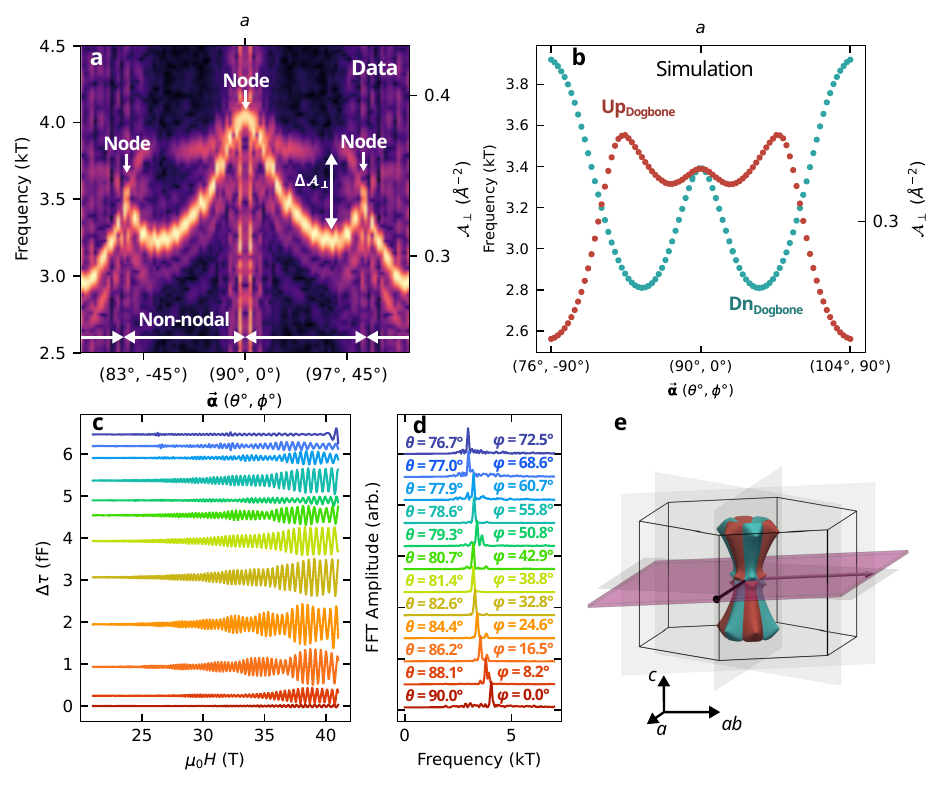}
\end{center}
\caption{\textbf{Confirmation of $g$-wave altermagnetic spin texture in CrSb. a,} Heatmap of quantum oscillation frequencies versus rotation angle (data have been symmetrised for ease of presentation) and \textbf{b,} simulated spin-up (red) and spin-down (blue) angular frequency profile computed by DFT. Here we rotate \textbf{H} through a plane of low symmetry at a tilt of 14$\degree$ from the $ab$ plane, which is depicted in Extended Data Fig. 4. We parameterise the orientation of \textbf{H} in the reference frame of the rotator mechanism by $\alpha$ that defines the angle within the rotation plane, which can be mapped back into the reference frame of the crystal by $\cos(\theta) = \sin(14\degree)\sin(\alpha)$ and $\tan(\varphi) = \cos(14\degree)\tan(\alpha)$. Symmetry-enforced nodal planes are crossed at $\varphi = 0\degree$ and at $\varphi = \pm 60\degree$. At all other orientations in this low-symmetry plane, significant spin-splitting is expected by DFT, and observed experimentally. \textbf{c,d,} Selected $\Delta \tau$ curves and corresponding FFTs that contributed to the heatplot in \textbf{a} (full dataset in Extended Data Fig. 5). \textbf{e,} The tilted rotation plane through which \textbf{H} was swept, whereby the arc that is traced out crosses a nodal plane every 60$\degree$ increment of $\varphi$, due to the $g$-wave symmetry profile.}
\label{Fig3}
\end{figure}

To trace out the spin-splitting of the CrSb Fermi surface away from its nodal planes, we measured QOs in a tilted plane of rotation, chosen such that an arc of low symmetry connects two symmetry-enforced nodal orientations. Along this arc, the spin-splitting should be considerable, but collapse to yield monofrequency waveforms at nodal orientations. We experimentally performed this by first rotating the sample platform 14$\degree$ orthogonally to the axis of the solenoid before loading it into the magnet. Then we rotated \textit{in-situ} through a polar angle (labelled $\alpha$ in Fig.~\ref{Fig3}), about an axis orthogonal to \textbf{H}, but always at a tilt of 14$\degree$ relative to the sample holder. This setup means that \textbf{H} may still be oriented along $a$, but subsequent rotation of the sample holder increments $\theta$ (between $c-a$) and $\varphi$ (between $a-ab$) such that $\cos(\theta) = \sin(14\degree)\sin(\alpha)$ and $\tan(\varphi) = \cos(14\degree)\tan(\alpha)$ (see Extended Data Fig. 4 for further details). We plot the resulting data for these rotational measurements in Fig.~\ref{Fig3}, including the simulated up- and down-sheet frequency profiles computed from DFT. 

The correspondence between experimental observation (Fig.~\ref{Fig3}a) and theoretical prediction (Fig.~\ref{Fig3}b) is very good. A singular frequency peak is observed for \textbf{H} $\parallel a$ ($\theta = 90\degree, \varphi = 0\degree$ in Fig.~\ref{Fig3}d), giving a bright spot at the nodal location of $\varphi = 0\degree$ in Fig.~\ref{Fig3}a. Upon incrementing $\alpha$, this then splits into two distinct branches, which, within resolution, come back together again close to $\varphi = \pm 60\degree$ as the next nodal plane is crossed, as expected for $\Delta(\vec k)$ possessing $g$-wave symmetry.

\begin{figure}[!htbp]
\vspace{-2.65cm}
\begin{center}
\includegraphics[width=0.9\linewidth]{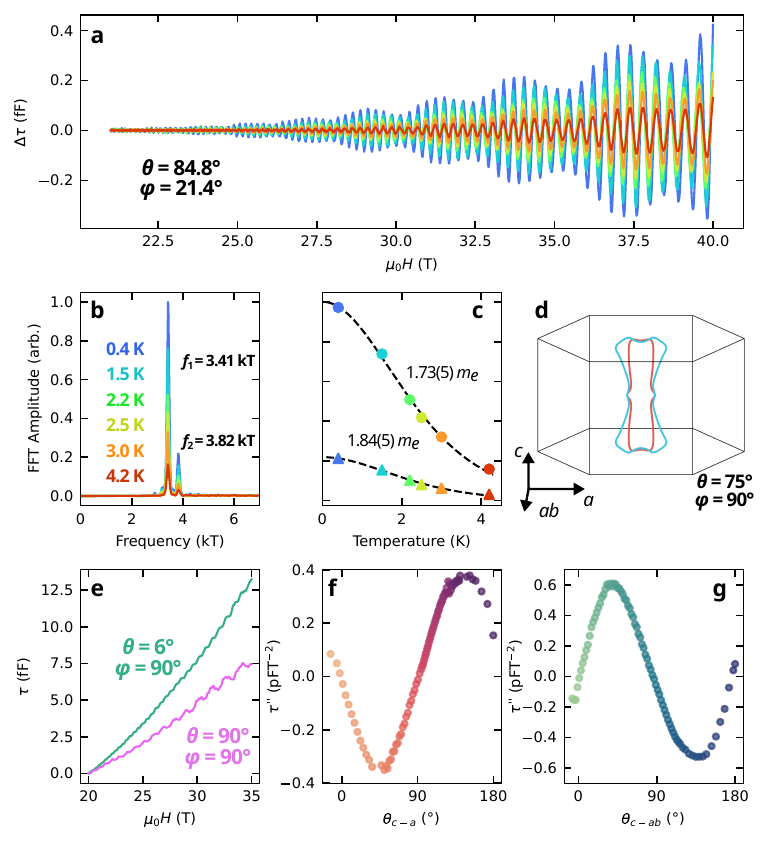}
\end{center}
\caption{\textbf{Spin-split effective mass study. a,} QOs at incremental temperatures for $\theta = 84.8\degree$, $\varphi = 21.4\degree$ and \textbf{b,} their corresponding frequency spectra, with temperatures indicated. $\Delta \tau$ was high-pass filtered to focus on two frequency components, at $f_1 =$ 3.41~kT and $f_2 =$ 3.82~kT (see \textit{Methods}), which come from spin-split up and down sheets of the dogbone pocket. \textbf{c,} Temperature dependence of QO amplitude fitted to the Lifshitz-Kosevich formula~\cite{LK}, yielding effective cyclotron masses of $m_1^* =$ 1.73(5) $m_e$ and $m_2^* =$ 1.84(5) $m_e$. The ratio of $\sqrt{\nicefrac{f_1}{f_2}}$ is equal to that of $\nicefrac{m_1^*}{m_2^*}$, as expected for these two frequency components being two spin-split daughters of the same mother Fermi sheet. \textbf{d,} Visualisation of spin-up and spin-down orbits around the dogbone cross-section, which can enclose markedly different areas for the same field orientation. \textbf{e,} Raw torque $\tau$ without background-subtraction at two orientations as indicated. Close to high-symmetry, the overall cantilever deflection is small, hence $\tau$ is dominated by the oscillatory component of magnetisation. \parfillskip=0pt}
\label{Fig4}
\end{figure}
\nopagebreak
\begin{figure}[!htbp]
    \ContinuedFloat
    \begin{centering}
    \caption{(cont.) The raw torque for $\theta=\varphi=90\degree$ has been scaled by a factor of 5 so that its magnitude is comparable to that at $\theta=6\degree, \varphi=90\degree$. \textbf{f,} Second derivative of torque with respect to field, $\tau''$, for rotations in the nodal $c-a$ plane and \textbf{g,} in the antinodal $c-ab$ plane. Although the QO frequency profiles for these two planes are markedly different, with the loss of mirror symmetry for each spin-split frequency trace in the antinodal plane (Fig.~\ref{Fig2}e), the background torque profile still exhibits the higher symmetry representative of the crystal structure.}
    \end{centering}
\end{figure}
%hpf 3000 T loess window 0.6 and poly 2

\subsection*{Altermagnetic scaling of $m_\uparrow^*/m_\downarrow^* = \sqrt{f_\uparrow/f_\downarrow}$}

We determined the effective cyclotron masses of the two spin-split surfaces in a Lifshitz-Kosevich temperature dependence study of QO amplitude in this rotation plane at $\alpha = 22\degree$, which in the crystal reference frame corresponds to \textbf{H} aligned at $\theta = 84.8\degree, \varphi = 21.4\degree$ (Fig.~\ref{Fig4}a). Two peaks are clearly resolved, with $f_1$ = 3.41~kT and $f_2$ = 3.82~kT. Fitting the temperature-dependent oscillatory damping to Lifshitz-Kosevich theory~\cite{LK} yields effective masses $m^*_1$ = 1.73(5)~$m_e$ and $m^*_2$ = 1.84(5)~$m_e$, where $m_e$ is the bare electron mass. For two QO frequency branches that are spin-split daughters of the same mother sheet, the daughters are expected to share the same mass renormalisation factors and Fermi velocity profile. The QO mass is given by $m^*=(\partial \mathcal{A_\perp}/\partial E) \hbar^2/(2\pi)$, where $\partial \mathcal{A_\perp}/\partial E = (\partial \mathcal{A_\perp}/\partial k)/(\hbar v_{\text{F}})$. If the orbit area scales with the square of the orbit size, $(\partial \mathcal{A_\perp}/\partial k) \propto \sqrt{\mathcal{A_\perp}}$, with $\mathcal{A_\perp}$ in turn $\propto$ the associated QO frequency. If $v_{\text{F}}(k)$ is the same around the two orbits, we would therefore expect $m_1^*/m_2^* = \sqrt{f_1/f_2}$. This is indeed what we observe: $m_1^*/m_2^* = 0.94\pm 0.04 \simeq \sqrt{f_1/f_2} = 0.945$. This result provides strong corroboration that these two frequencies may therefore be attributed to altermagnetic spin-splitting of the same parent sheet. The maximum spin-splitting can likewise be estimated from the frequency difference $\delta f = f_2 - f_1 = 0.41$~kT, which according to the Onsager relation~\cite{onsager_rel} translates to a difference between the areas enclosed by spin-up and spin-down orbits as $\delta A = \delta f (2\pi e)/\hbar = \Delta \partial A/\partial \varepsilon  = m^* \Delta (2\pi/\hbar^2)$, with $\Delta$ the energy splitting at the Fermi level. With the observed average effective mass $m^* \simeq 2.03 m_e$, we thereby estimate a spin-splitting at the Fermi level of $\Delta \approx 25$~meV for this field orientation.

\subsection*{$g$-wave magnetic order parameter}

Putting together our results from the three different rotation planes presented in Figs.~\ref{Fig2}~\&~\ref{Fig3}, we can firmly pin down the unconventional magnetic order parameter symmetry of CrSb. By tracking how the spin-split QO frequency profile of the primary CrSb Fermi pocket evolves under rotation, we identify four nodal planes in total: three defined for all $\theta$ at $\phi=0\degree$, $60\degree$, $120\degree$, and one for all $\phi$ at $\theta=90\degree$. We can therefore compare the symmetry of our unconventional magnetic order parameter, $\Delta(\theta,\phi)$, to that of a real spherical harmonic. We find that $\Delta(\theta,\phi)$ obeys the symmetries of $\mathcal{Y}_{4}^{-3}(\theta, \phi) \propto P_4^3(\cos\theta)\sin3\phi$ defined in terms of the associated Legendre polynomial $P_\ell^m(x)$~\cite{BLANCO199719} (see \textit{Methods}). This harmonic transforms as the $B_{1g}$ irreducible representation of the $D_{6h}$ point group of CrSb. $\mathcal{Y}_{4}^{-3}(\theta, \phi)$ can be expressed in Cartesian form as $yz(3x^2-y^2)$ or equivalently in spherical polars as $r^4\sin^3(\theta)\cos(\theta)\sin(3\phi)$ (sketched in Fig.~\ref{Fig1}f). This function has nodes when $\theta = 0\degree, 90\degree$ and $\phi =0\degree$, $60\degree$, $120\degree$, just as $\Delta(\theta,\phi)$ does. Our study thereby demonstrates the sensitivity of QO measurements to spin-split band structures, positioning such measurements as an ideally suited diagnostic tool for ascertaining the order parameter symmetry of unconventional metallic magnets.

In summary, our systematic bulk-sensitive QO measurements have mapped out the unconventional magnetic order parameter symmetry of CrSb, revealing the $g$-wave altermagnetic spin texture. This manifests in the $k$-dependence of the spin-splitting per Fermi pocket, $\Delta(\vec k)$. For magnetic fields applied in highly symmetric nodal planes, $\Delta(\vec k)$ is always zero -- whereas, rotating through arcs of low symmetry $\Delta(\vec k)$ is generally large and well-resolvable, collapsing back to zero upon crossing through a nodal plane. For fields applied in nodal planes, quasiparticle orbits are spin-degenerate and mirror-symmetric. By contrast, in antinodal planes this mirror symmetry is lost, leading to spin-split QO frequency versus angle traces -- providing a hallmark signature of altermagnetic ordering. Additionally, we observe excellent quantitive agreement in the ratios between quantum oscillatory frequencies and effective masses, as theoretically expected for spin-split Fermi surface sheets. In combination, these results provide strong empirical evidence supporting the designation of CrSb as a room-temperature metallic altermagnet. Furthermore, our optimised growth procedure has produced high-quality crystals with residual resistivities as low as 2~$\upmu \Omega$cm. Given that CrSb is relatively insensitive to air and moisture, it can be synthesised to high purity, the altermagnetic ordering temperature is very high, and both chromium and antimony are relatively abundant easily-sourced elements, this combination of favourable properties positions CrSb as a leading material for pursuing applications in next-generation low-energy spintronic devices.

%We observed a profile of QO frequency versus rotation angle that obeys the mirror-symmetric $C_{2v}$ point group, consistent with the crystal structure when rotating the orientation of a magnetic field through nodal planes. By contrast, for rotations in an antinodal plane, this lowers to $C_2$ symmetry per non-degenerate spin-split Fermi surface sheet, losing mirror-symmetry -- a smoking-gun signature of altermagnetic ordering. We observe significant spin splitting upon applying a magnetic field in low-symmetry orientations, resolved as two distinct frequency traces, which come together at highly symmetric nodal orientations due to symmetry-enforced Kramers degeneracy. These results thereby firmly establish CrSb as a room-temperature metallic altermagnet. Our optimised growth procedure has produced high-quality crystals with residual resistivities as low as 2~$\upmu \Omega$cm, making CrSb a promising candidate for exploring next-generation low-energy spintronic device applications.

% \begin{figure}[t!]
% \vspace{-0cm}
% \begin{center}
% \includegraphics[width=1\linewidth]{}
% \end{center}
% \caption{\textbf{xx a,} words}
% \label{fig:tuning}
% \end{figure}

\clearpage
\normalsize{
\bibliographystyle{naturemag_noURL}
\bibliography{CrSb}
}

\clearpage
\Large
\noindent
\textbf{Methods}

\normalsize
\vspace{5mm}\noindent 
\textbf{Crystal growth}\\ Single crystal CrSb specimens were grown by the chemical vapour transport technique. Stoichiometric amounts of Cr (chunks, 99.995$\%$) and Sb (Shots, 99.9999$\%$) were used as source material. Iodine was added as a transport agent, calculated to have a pressure of 1 bar at growth conditions. The starting materials were sealed under vacuum in a quartz ampoule and placed in a horizontal two-zone furnace. The temperature was slowly ramped up to $T_1$ = 925 °C and $T_2$ = 900 °C, left for two weeks, and subsequently cooled at the furnace cooling rate to room temperature. The resulting crystals were hexagonal platelets up to 1.5 mm in diameter, along with larger areas possessing intergrown crystals of CrSb, several mm in size. Only single-crystal specimens were used in this study.

\vspace{5mm}\noindent 
\textbf{Sample characterisation}\\ Several crystals were picked from a batch of single crystals and crushed into a fine powder. This powdered sample was then distributed on a microscope slide, which had a thin layer of vacuum grease. Powder X-ray diffraction was measured in the Bragg-Brentano geometry on a Bruker D8, using a Cu-source, with the results plotted in Extended Data Fig. 1. The measurement was performed in a 2$\theta$ range of 10$^\circ$ - 90$^\circ$, with no peaks observed below 20$^\circ$.

The obtained data display sharp, well-defined peaks, indicating a high level of crystallinity. The data were analysed using the Rietveld method, yielding an excellent fit (R$_\text{Bragg}$ = 3.39), which describes all observed peaks, thereby indicating that the samples are phase pure. The measured crystal structure is in good agreement with prior studies~\cite{willis1953crystal}.

We also performed electrical transport, magnetisation, and Laue diffractometry measurements (Extended Data Fig. 1). Samples were predominantly screened by temperature-dependent resistivity measurements, utilised to extract their residual resistivity ratios (RRRs). To do this, we fitted the low-temperature data to the square of the temperature and extrapolated to absolute zero to determine the residual resistivity. The 300 K resistivity was then divided by this value to yield the RRR. Higher RRR values indicate longer mean free paths and hence higher crystalline quality. Typical RRR values were in the approximate range of 10-28. High-quality specimens were then oriented by Laue diffractometry, in preparation for high magnetic field dHvA measurements.

\vspace{5mm}\noindent 
\textbf{de Haas-van Alphen effect torque magnetometry measurements}\\ High-quality samples were selected following characterisation screening and brought to the National High Magnetic Field Laboratory (NHMFL), Tallahassee, Florida, USA. For torque magnetometry measurements, we largely followed the methodology outlined in ref.~\cite{Eaton2024}. Samples were mounted on flexible BeCu cantilevers and affixed using multiple layers of General Electric low-temperature varnish, giving good thermal contact and strong adhesion between sample and cantilever. Cantilevers were soldered in place, such that the cantilever head was suspended above a copper baseplate, by a short separation distance. As the magnetic field was swept, the change in capacitance between the cantilever and baseplate, due to the magnetic torque exerted on the sample, was measured by a General Radio analogue capacitance bridge using phase-sensitive detection. The change in torque was calibrated to units of Farads using an Andeen Hagerling digital capacitance bridge.

All dHvA measurements were performed in the 41.5~T all-resistive magnet in Tallahassee. A $^3$He sample environment was utilised, along with a probe-mounting of our custom design. Rotations of the sample orientation with respect to the magnetic field were performed \textit{in-situ} utilising a brushless linear motor. Angles were calibrated by the change in sign of the torque background -- identifying high symmetry directions of the crystal -- and verified using a Hall sensor.

The oscillatory component $\Delta \tau$ was isolated from the background magnetic torque $\tau$ by performing a LOESS~\cite{LOESS} subtraction. In general, due to the intricate web sheet of the CrSb Fermi surface, the dHvA waveform at a given angle could be quite complicated due to the presence of numerous frequency components. To simplify our analysis and concentrate on the dogbone Fermi sheet, we often performed combined high-pass filtering with short LOESS windows in our analysis. The dogbone frequencies are most prominent above 3~kT, and so we Butterworth high-pass filtered frequencies in inverse field in this range. This was combined with a short sliding LOESS window over $\tau$, which effectively fits any slow oscillations in with the background (assumed to be quadratic in $H$), therefore producing a $\Delta \tau$ waveform dominated by higher frequency components. For the $\Delta \tau$ traces presented in Fig.~\ref{Fig1}, this involved using a LOESS window of 0.7~T. In Fig.~\ref{Fig2}, we used a window of length 1.2~T to show the strong spectral weight at lower frequencies due to the web. By contrast, in Fig.~\ref{Fig4} we utilised a window of only 0.6~T to focus on the $>$~3~kT components in our temperature dependence study.

%\newpage
\vspace{5mm}\noindent 
\textbf{Density functional theory calculations}\\ 
Density functional theory (DFT) calculations for CrSb were performed using the all-electron, full-potential linearised augmented plane-wave (FP-LAPW) method as implemented in the \texttt{WIEN2k} code~\cite{wien2k}. The electronic structure was converged on a $43 \times 43 \times 28$ Monkhorst--Pack $k$-point mesh within the Brillouin zone of the primitive hexagonal unit cell. Exchange--correlation effects were treated within the Generalised Gradient Approximation (GGA). We specified two distinct Cr sites (Cr$_1$ and Cr$_2$) within the primitive unit cell, corresponding to Cr atoms adopting up and down spin polarisation. Calculations were initialised so that one Cr site has a higher spin-up density and the other has a spin-down density. The onsite spin-polarisation was then allowed to vary throughout the self-consistency cycles until the compensated collinear ground state was reached. Quantum oscillation frequency analysis of the resultant Fermi surface sheets was determined using \texttt{SKEAF}~\cite{skeaf}. Fermi surface visualisation was performed using \texttt{py\_FS}~\cite{py_FS, Eaton2024}.

We assumed that ambient pressure CrSb in the NiAs‐type structure (P6$_3/mmc$) adopts lattice parameters $a = 4.12$~\AA, $b = 4.12$~\AA,  and $c=5.47$~\AA. Within the unit cell, there are two equivalent Cr sites:
\begin{table}[H]
\begin{center}
\def\arraystretch{1.5}
\centering
\begin{tabular}{ cccc }
\toprule
Atom & X & Y & Z \\
\midrule 
\midrule
Cr$_1$ & 0.00 & 0.00 & 0.00 \\
Cr$_2$ & 0.00 & 0.00 & 0.50 \\
\bottomrule
\end{tabular}
\end{center}
\end{table}
\noindent
and two equivalent Sb sites: 
\begin{table}[H]
\begin{center}
\def\arraystretch{1.5}
\centering
\begin{tabular}{ cccc }
\toprule
Atom & X & Y & Z \\
\midrule 
\midrule
Sb$_1$ & 0.67 & 0.33 & 0.25 \\
Sb$_2$ & 0.33 & 0.67 & 0.75 \\
\bottomrule
\end{tabular}
\end{center}
\end{table}
\noindent We reduce the symmetry of the crystal lattice from P6$_3/mmc$ to P3$m$1 by specifying that the two Cr sites adopt opposite spins.

DFT calculations converge on a Fermi surface where the bands associated with the down and up `dogbone' surfaces are open about the A high symmetry point, corresponding to a cylindrical topology. This is inconsistent with our quantum oscillation measurements, where we resolve oscillations from these sheets for magnetic fields very close to the $a$ and $ab$ directions. No frequencies would be observed for these field orientations if the sheets were cylindrical. Therefore, we propose that these bands form closed Fermi surface sheets with dogbone-like geometry. To `close' the open Fermi surface sheets of our DFT calculations, we shift our band edges relative to the Fermi energy. The dogbone-like sheets were shifted down by 0.11~eV so that the calculated frequencies along the $a$, $ab$, and $c$-directions are in good agreement with the quantum oscillation data. Since the dogbone sheets are of hole-character, we shifted up the `web' sheets (of electron-character) by 0.015~eV to keep the total carrier number constant.

\vspace{5mm}\noindent 
\textbf{Consideration of spin-orbit coupling} \\
Real materials always exhibit some spin-orbit coupling (SOC) and many-body electronic correlations, meaning a purely non-relativistic framework is only ever an idealisation. Nevertheless, our symmetry-based interpretation remains robust. Because CrSb possesses an inversion-symmetric crystal structure, the spatial symmetries protecting the orientation of the nodal planes remain intact. Furthermore, under the intense magnetic fields utilised in our experiments, field-assisted tunnelling (magnetic breakdown) allows quasiparticles to traverse small hybridisation gaps opened by weak SOC, effectively restoring the pristine altermagnetic trajectories. As detailed in the \textit{Supplementary Information} (in which we also explicitly account for electronic correlations) these considerations justify our simplified symmetry picture introduced here, yielding a direct mapping between the quantum oscillation frequency spectra and the underlying altermagnetic order parameter $\Delta(\vec k)$.

\vspace{5mm}\noindent 
\textbf{Energy Splitting from Quantum Oscillation Frequencies} \\
From the Onsager relation~\cite{onsager_rel}, we can equate a quantum oscillation frequency to a reciprocal space area as 
\begin{equation}
\label{eq:onsager}
    F(E) = \frac{\hbar}{2\pi e}A(E)
\end{equation}
\noindent The cyclotron mass of an orbit, $m_c$, is related to the rate of change of the orbital area by
\begin{equation}
    m_c = \frac{\hbar^2}{2\pi} \frac{\partial A}{\partial E}\bigg\rvert_{E_f}
\end{equation}
By taking the derivative of Eq.~\ref{eq:onsager} with respect to $E$ we then determine
\begin{equation}
    \frac{dF}{dE} =\frac{\hbar}{2\pi e}\frac{\partial A}{\partial E} = \frac{\hbar}{2\pi e} \frac{2\pi}{\hbar^2}m_c = \frac{m_c}{e\hbar}
\end{equation}
\noindent We can use this to determine the energy difference associated with the frequency splitting of two bands
\begin{equation}
    \Delta E \sim \Delta F \frac{dE}{dF} = \frac{e\hbar}{m_c} \Delta F
\end{equation}

\vspace{5mm}\noindent 
\textbf{Spherical Harmonic Notation}\\ 

In the text, we represent the symmetry of the altermagnetic spin-splitting of CrSb in terms of the real spherical harmonic $\mathcal{Y}_{4}^{-3}(\theta, \phi)$.

The complex spherical harmonics can be defined in terms of the associated Legendre Polynomials as $Y_\ell^m(\theta,\phi) = N_{\ell m} e^{im\phi}P_\ell^m(\cos\theta)$, where $N_{\ell m}$ is a normalisation factor, $P_\ell^m(x)$ is an associated Legendre polynomial, and $Y_\ell^m(\theta,\phi)$ is the complex spherical harmonic for $\ell \geq 0$ and $m \in [-\ell,\ell]$. The complex spherical harmonics are eigenfunctions of the total angular momentum operator $\hat{L}^2$ and of the generator of rotations about the azimuthal axis $\hat{L}_z$, spanning a complete orthonormal basis.

The complex spherical harmonics are defined up to a phase factor $e^{im\phi}$, and so their magnitude does not change as a function of $\phi$. Therefore, it is convenient to work in the basis of the real spherical harmonics, which have explicit $\phi$ dependence, when describing the symmetry of an unconventional magnetic order parameter. We can define the real spherical harmonics $\mathcal{Y}_{\ell}^m(\theta,\phi)$ in terms of linear combinations of complex harmonics according to:

\begin{equation}
\mathcal{Y}_{\ell}^m = \begin{cases} 
\frac{1}{\sqrt{2}} \left( Y_{\ell}^{-m} + (-1)^m Y_{\ell}^m \right) & \text{if } m > 0 \\
Y_{\ell}^0 & \text{if } m = 0 \\
\frac{i}{\sqrt{2}} \left( Y_{\ell}^{-|m|} - (-1)^{|m|} Y_{\ell}^{|m|} \right) & \text{if } m < 0,
\end{cases}
\end{equation}
or equivalently, in terms of the associated Legendre polynomials:
\begin{equation}
    \mathcal{Y}_{\ell}^m = 
    \begin{cases} 
        \sqrt{2} (-1)^m N_{\ell m} P_\ell^m(\cos \theta) \cos(m\phi) & \text{if } m > 0 \\
        N_{\ell 0} P_\ell^0(\cos \theta) & \text{if } m = 0 \\
        \sqrt{2} (-1)^m N_{\ell |m|} P_\ell^{|m|}(\cos \theta) \sin(|m|\phi) & \text{if } m < 0. 
    \end{cases}
\end{equation}

Defining the real spherical harmonics this way means they form a complete set that spans the same basis as the complex spherical harmonics; however, importantly, they have well-defined varying magnitudes as a function of $\phi$. This allows us to map the $\mathcal{Y}_{4}^{-3}$ real spherical harmonic to the $g$-wave symmetry profile of the altermagnetic order parameter in CrSb.

\vspace{5mm}\noindent 
\textbf{Contactless resistivity measurements}\\ 

Contactless resistivity measurements were conducted using the proximity detector oscillator~\cite{PDO_Altarawneh} (PDO) technique. A selected CrSb sample was mounted on a hand-wound planar coil of 15 turns, acting as the inductive component of the oscillator. The coil diameter was customised to match the sample width for optimal filling factor. A counter-wound outer coil enclosing the same area as the inner coil was added to compensate magnetic flux induced during the field pulse, minimising background pickup.

As the applied magnetic field is swept, changes in the sample's resistivity $\rho$ and susceptibility $\chi_{s}$ lead to changes in the inductance of the oscillator and produce a shift in the resonant frequency of the oscillator, which can be described by 
\begin{equation}
\frac{\Delta f}{f} \approx -\,\eta\,\frac{\delta}{d}
\left(
\mu_r \frac{\Delta \rho}{\rho} + \Delta \chi_s
\right),
\label{eq:pdo_df_over_f}
\end{equation}
where $\eta$ is the filling factor, $d$ is the sample thickness, and $\mu_r = 1+\chi_s$ is the relative magnetic permeability~\cite{PDO_Altarawneh}. For a metallic material like CrSb, eddy current restricts the penetration of the RF field to a characteristic skin depth $\delta = \sqrt{2\rho/(\mu_r \mu_0 \omega)}$, where $\omega$ is the excitation frequency, such that the frequency response is dominated by changes in the resistivity $\rho$.

PDO measurements reported in this study were performed in a 65~T pulsed magnet at the Dresden High Magnetic Field Laboratory (HLD) in Dresden, Germany, following the methodology of ref.~\cite{tony2024enhanced}. A customised $^3$He cryostat was fitted to the magnet, providing a base temperature of $\sim$600~mK throughout the pulses. A raw resonant frequency $\sim$25~MHz was achieved, followed by a heterodyne mixing circuit to down-convert the signal to $\sim$10.5~MHz, which was subsequently acquired using a high definition oscilloscope.

Quantum oscillatory components were analysed over a magnetic field range of 38 – 63~T using a LOESS background subtraction with an 8~T window and a second-order polynomial background subtraction. A quantum oscillation of frequency 0.8~kT was clearly resolved (Extended Data Fig. 7).

We note that, for sufficiently high magnetic fields, altermagnets have been predicted to exhibit certain distinguishing quantum oscillatory features, such as a distinct frequency-splitting at a field-induced Lifshitz transition separating the up and down sheets~\cite{li2024diagnosingaltermagneticphasesquantum}. However, we measured up to a maximal field strength of 64~T (see Extended Data Fig.~7), and observed no such signatures. This is likely due to the very high ordering temperature (and hence energy scale) of altermagnetism in CrSb, which remains robust up to these large field strengths.

%\clearpage

\Large
\vspace{12mm}\noindent
\textbf{Acknowledgements}

\normalsize
\vspace{5mm}\noindent 
We are grateful to A.J. Hickey, A. Agarwal, T.L. Lefebvre, J. Knolle, D. Candido, G.G. Lonzarich and N.J.M. Popiel for stimulating discussions. We thank T. Haidamak, M. Vali\v{s}ka, H. Li, P. Einarsson Nielsen, H. Chen, R. Mann, T.J. Brumm, and especially D. Svit\'{a}k and S. Hayden, for technical advice and assistance. We appreciate creative input from H. Weinberger.

\Large
\vspace{12mm}\noindent
\textbf{Funding statement}

\normalsize
\vspace{5mm}\noindent 
This project was supported by the EPSRC of the UK through grants EP/Z533695/1 and EP/R513180/1. A portion of this work was performed at the National High Magnetic Field Laboratory, which is supported by National Science Foundation Cooperative Agreement Nos. DMR-1644779 \& DMR-2128556 and the State of Florida. We acknowledge support of the HLD at HZDR, a member of the European Magnetic Field Laboratory (EMFL). T.I.W. and A.G.E. acknowledge support from ICAM through US National Science Foundation (NSF) Grant Number 2201516 under the Accelnet program of the Office of International Science and Engineering and from QuantEmX grants from ICAM and the Gordon and Betty Moore Foundation through Grant GBMF9616. T.I.W. acknowledges support from Murray Edwards College (University of Cambridge) and the Cambridge Philosophical Society through a Henslow Fellowship. A.G.E. acknowledges support from Sidney Sussex College (University of Cambridge).

\Large
\vspace{12mm}\noindent
\textbf{Author contribution statement}

\normalsize
\vspace{5mm}\noindent 
M.L., M.F.H. and R.T. grew single crystal CrSb specimens. M.L. characterised samples. M.L., T.I.W., Z.W., M.S., D.G., Y.S. and A.G.E. performed quantum oscillation experiments. M.L., T.I.W., Z.W., F.M.G. and A.G.E. analysed and interpreted data. T.I.W. performed density functional theory calculations. A.G.E. conceived the project. F.M.G. and A.G.E. supervised the project. T.I.W., F.M.G. and A.G.E. wrote the paper, with input from all co-authors.

\vspace{12mm}
\Large
\noindent
\textbf{Competing interests statement}

\normalsize
\vspace{5mm}\noindent 
The authors declare no competing interests.

\vspace{12mm}
\Large
\noindent
\textbf{Data availability}

\normalsize
\vspace{5mm}\noindent 
The datasets supporting the findings of this study are available from the University of Cambridge Apollo Repository~\cite{CrSbFSdata}.

%\clearpage

\clearpage
% \begin{figure}[t!]
%     \vspace{2cm}
%     \begin{center}
%     \includegraphics[width=.8\linewidth]{Extended/Torque.pdf} 
%     \end{center}
%     \vspace{-5pt}
%     \caption*{\textbf{Cartoon|} }
%     \label{pic:14T_tuning}
% \end{figure}

\begin{figure}[t!]
    \vspace{-2cm}
    \begin{center}
    \includegraphics[width=.9\linewidth]{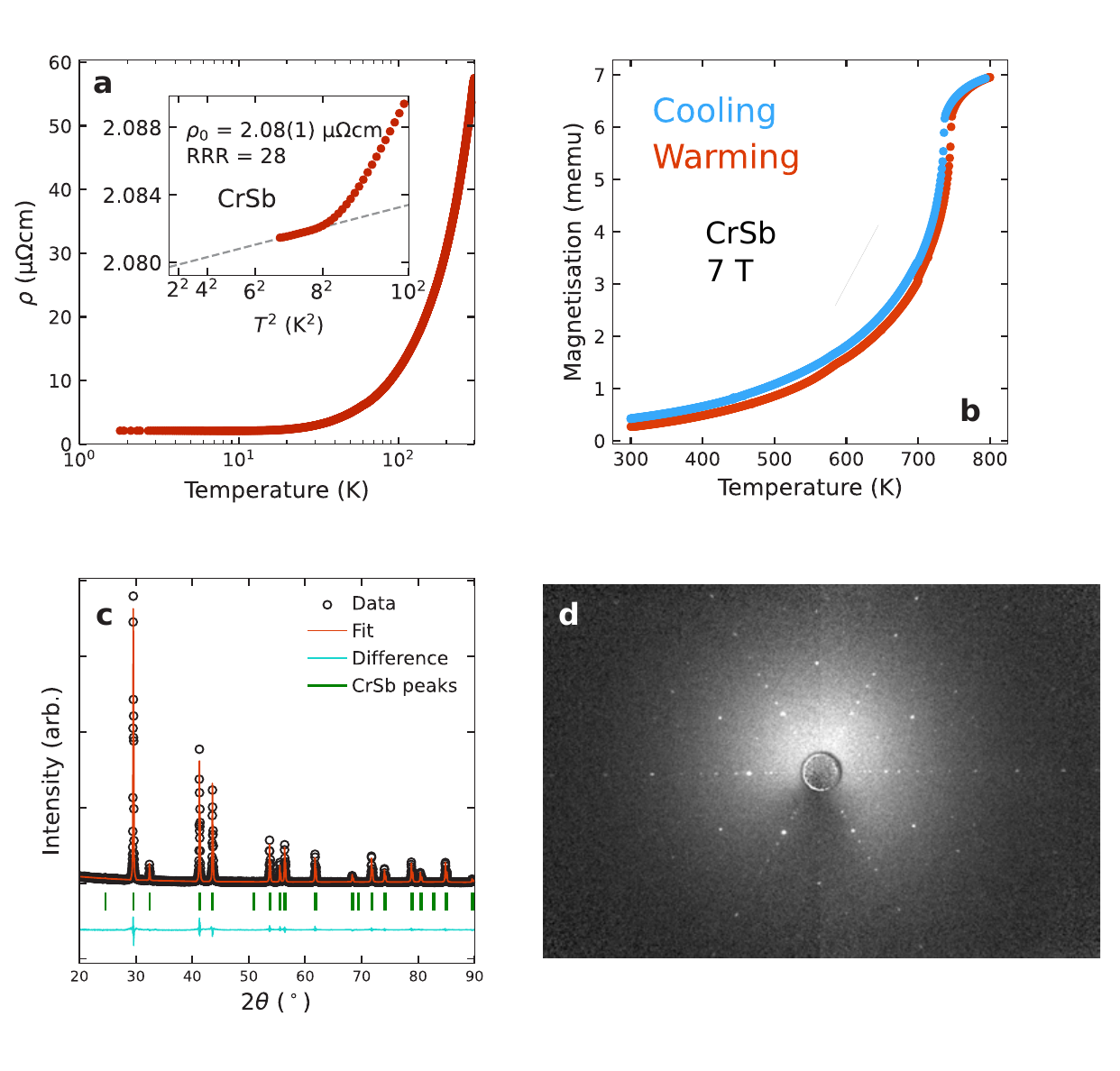}
    \end{center}
    \vspace{-5pt}
    \caption*{\textbf{Extended Data Fig. 1 | Sample characterisation studies. a,} Electrical resistivity $\rho$ of CrSb measured by the four-terminal technique. The inset shows low temperature data fitted quadratically in temperature, to extract the residual resistivity of 2.08(1) $\upmu \Omega$cm. This yields a residual resistivity ratio, upon dividing the 300~K resistivity by the fitted extrapolation to absolute zero, of RRR = 28. \textbf{b,} Magnetisation versus temperature up to 800~K, measured in a Quantum Design Magnetic Properties Measurement System with the furnace option mode under an applied field of 7~T. Red points were recorded on warming, with blue taken subsequently upon cooling. An anomaly at $\approx$~740~K is resolved, indicating the onset of compensated collinear magnetic order. This is slightly higher than previously reported~\cite{Snow52PhysRev.85.365}, likely due to improved crystal quality. \textbf{c,} Powder x-ray diffraction data plotted versus 2$\theta$. The recorded data agree very well with the known structure~\cite{willis1953crystal} of CrSb. \textbf{d,} Lauegram of a single crystal specimen utilised in our quantum oscillation study. Sharp spots in a hexagonal pattern are clearly resolved, confirming single-crystallinity.}
    \label{pic:14T_tuning}
\end{figure}

%\clearpage

\begin{figure}[t!]
    \vspace{-2cm}
    \begin{center}
    \includegraphics[width=.9\linewidth]{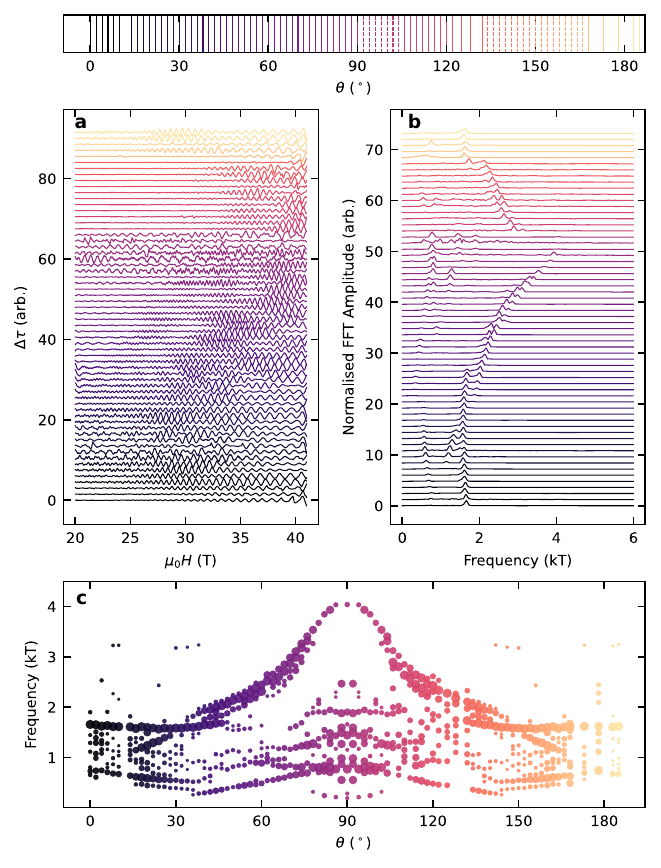} 
    \end{center}
    \vspace{-5pt}
    \caption*{\textbf{Extended Data Fig. 2 | Quantum oscillation measurements in the nodal $c-a$ rotation plane. a,} Normalised oscillatory component of magnetic torque $\Delta \tau$, \textbf{b,} the corresponding fast Fourier transform spectra, and \textbf{c,} the frequency versus angle profile. A magnetic field range of 20-41T was used, along with a LOESS window of 1.2~T and a second-order polynomial (see \textit{Methods}). Dashed lines in the angular scale bar denote where symmetry-equivalent traces have been reflected through $\theta =$ 90$\degree$, for display purposes in panel \textbf{c}. No spin-splitting is resolved in this nodal plane.}
    \label{pic:14T_tuning}
\end{figure}

\begin{figure}[t!]
    \vspace{-2cm}
    \begin{center}
    \includegraphics[width=.9\linewidth]{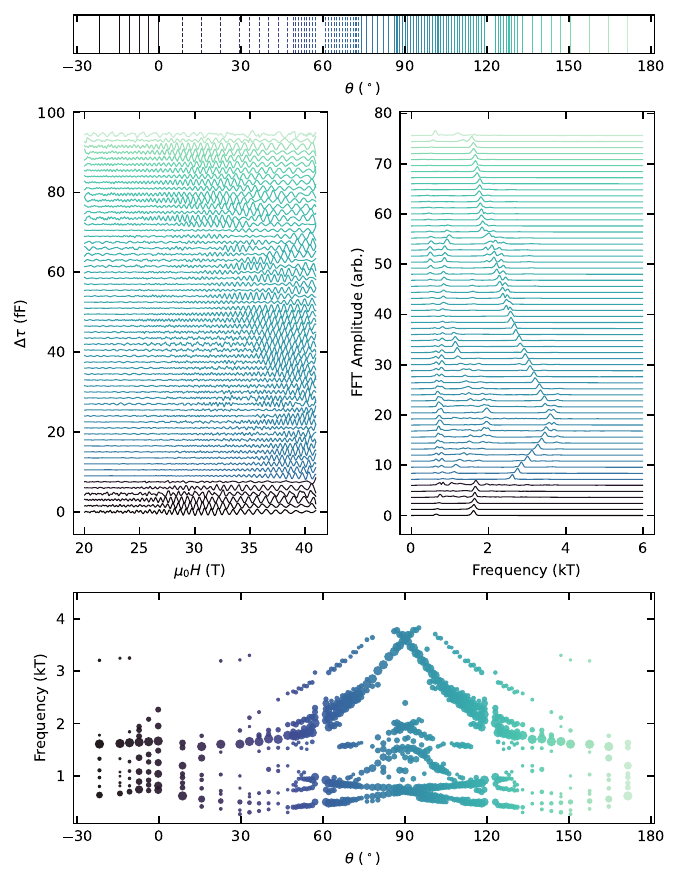} 
    \end{center}
    \vspace{-5pt}
    \caption*{\textbf{Extended Data Fig. 3 | Quantum oscillation measurements in the antinodal $c-ab$ rotation plane. a,} Normalised oscillatory component of magnetic torque $\Delta \tau$, \textbf{b,} the corresponding fast Fourier transform spectra, and \textbf{c,} the frequency versus angle profile. A magnetic field range of 20-41T was used, along with a LOESS window of 1.2~T and a second-order polynomial (see \textit{Methods}). Dashed lines in the angular scale bar denote where symmetry-equivalent traces have been reflected through $\theta =$ 90$\degree$, for display purposes in panel \textbf{c}. Significant spin-splitting is resolved close to $\theta =$ 90$\degree$, as shown clearly in Fig.~\ref{Fig2}.}
    \label{pic:14T_tuning}
\end{figure}

\begin{figure}[t!]
    \vspace{-2cm}
    \begin{center}
    \includegraphics[width=\linewidth]{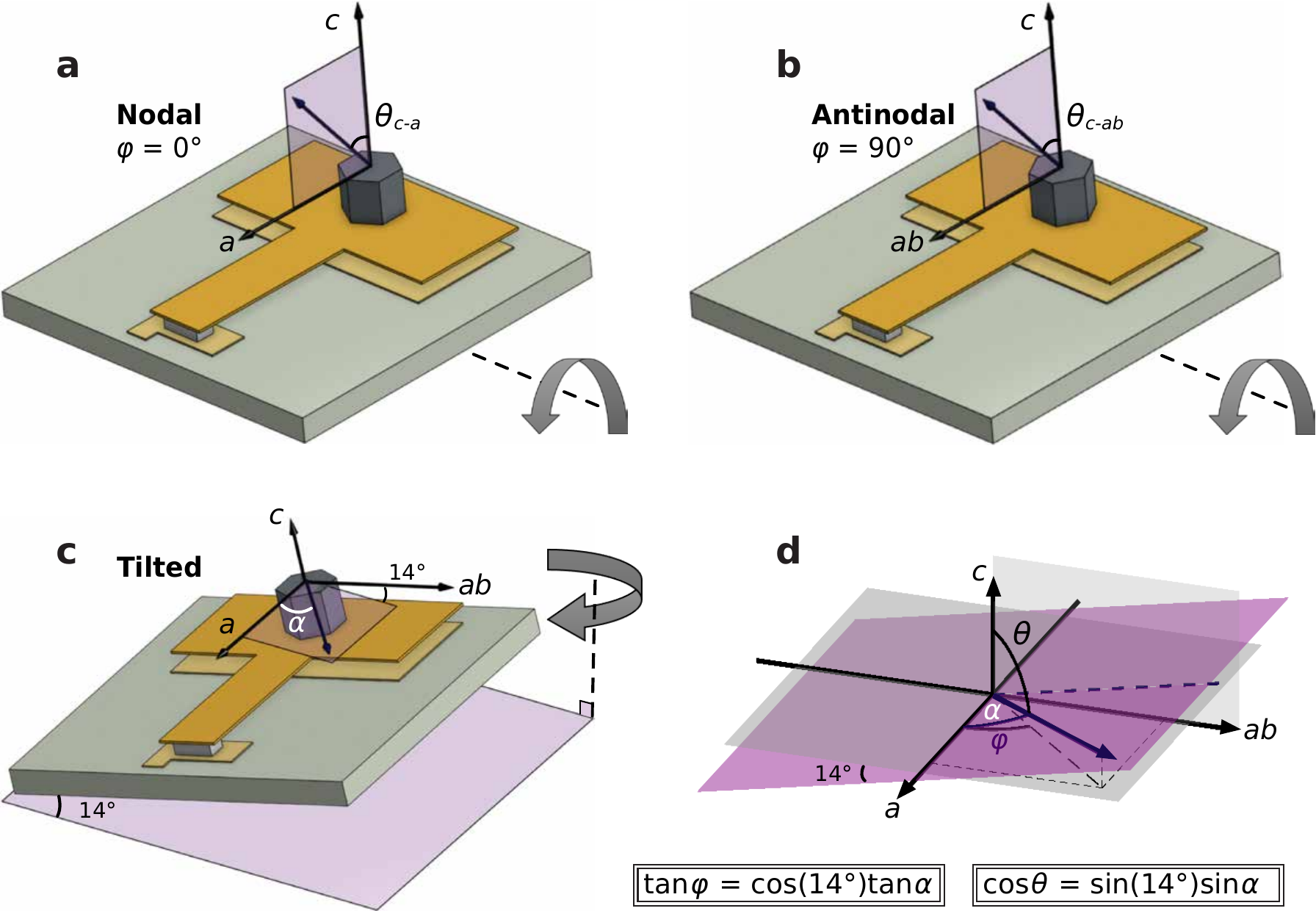} 
    \end{center}
    \vspace{-5pt}
    \caption*{\textbf{Extended Data Fig. 4 | Schematic diagrams of the rotational planes measured in this study. a,} The relative crystal orientation with respect to the rotation axis for the nodal $c-a$ rotation plane and \textbf{b,} the same for the antinodal $c-ab$ plane. The dark grey hexagonal prism depicts the sample, which is mounted on a metallic cantilever narrowly suspended above a copper baseplate, therefore enabling capacitive torque magnetometry measurements. \textbf{c,} The tilted rotation plane for which data are presented in Fig.~\ref{Fig3}, for which the relative orientations of $\theta$ and $\varphi$ with respect to $\alpha$ are defined in \textbf{d}.}
    \label{pic:14T_tuning}
\end{figure}

\begin{figure}[t!]
    \vspace{-2cm}
    \begin{center}
    \includegraphics[width=.9\linewidth]{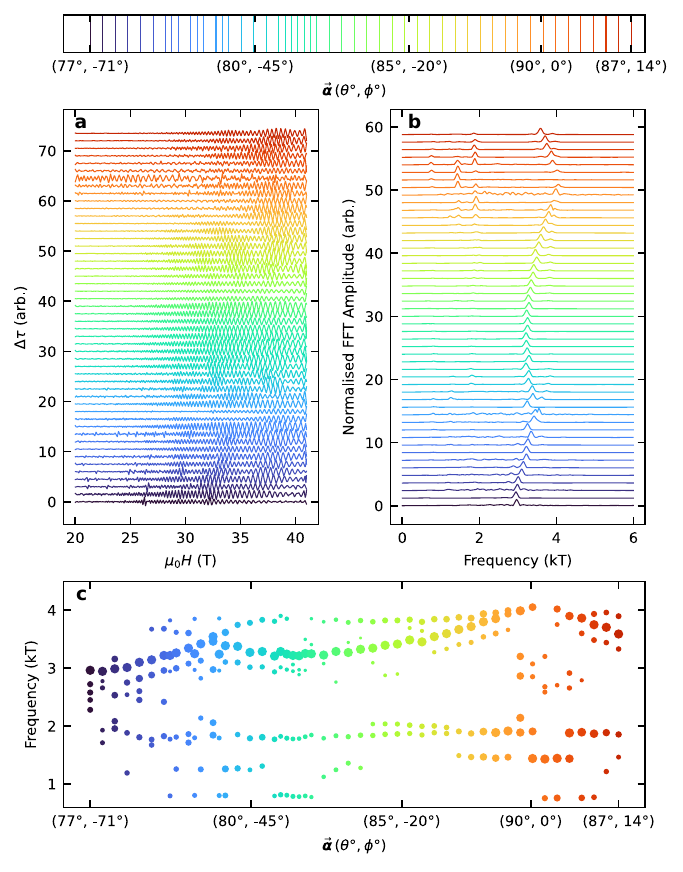} 
    \end{center}
    \vspace{-5pt}
    \caption*{\textbf{Extended Data Fig. 5 | Quantum oscillation measurements in a low-symmetry tilted rotation plane. a,} Normalised oscillatory component of magnetic torque $\Delta \tau$, \textbf{b,} the corresponding fast Fourier transform spectra, and \textbf{c,} the frequency versus angle profile. A magnetic field range of 20-41T was used, along with a LOESS window of 0.6~T and a second-order polynomial (see \textit{Methods}). The rotation plane is defined in Extended Data Fig. 4. The high frequency branch undergoes significant spin-splitting away from nodal planes, as shown clearly in Fig.~\ref{Fig3}. }
    \label{pic:14T_tuning}
\end{figure}

\begin{figure}[t!]
    \vspace{-2cm}
    \begin{center}
    \includegraphics[width=0.8\linewidth]{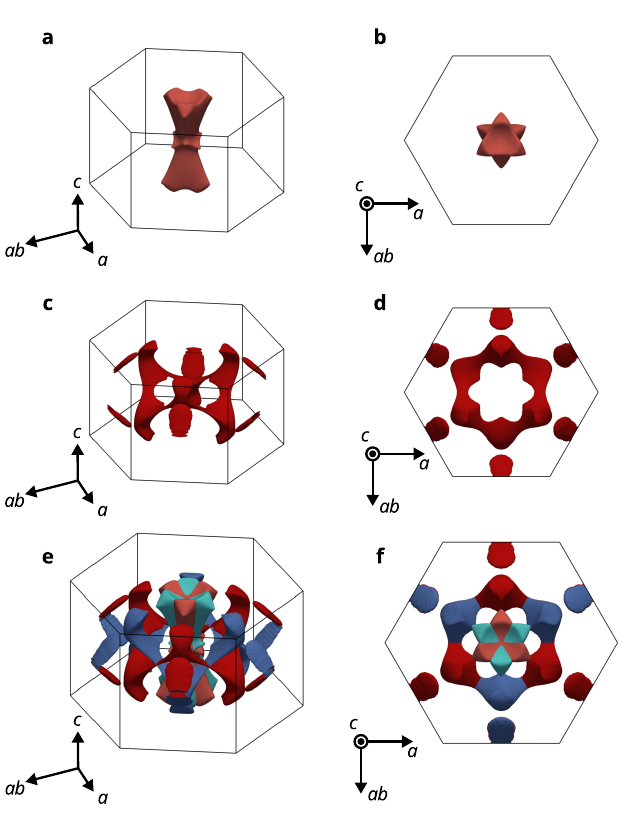} 
    \end{center}
    \vspace{-5pt}
    \caption*{\textbf{Extended Data Fig. 6 | Fermi surface renderings. a,b,} The central (dogbone) Fermi sheet due to two hole-type bands that cross the Fermi level, and \textbf{c,d,} the secondary (web) sheet that results from two electron-type sheets crossing the Fermi level. For clarity, we plot only the up-sheets here. \textbf{e,f} Spin-up (red) and spin-down (blue) sheets of the entire Fermi surface of CrSb.}
    \label{pic:big_fs}
\end{figure}

\begin{figure}[t!]
    \vspace{-2cm}
    \begin{center}
    \includegraphics[width=.8\linewidth]{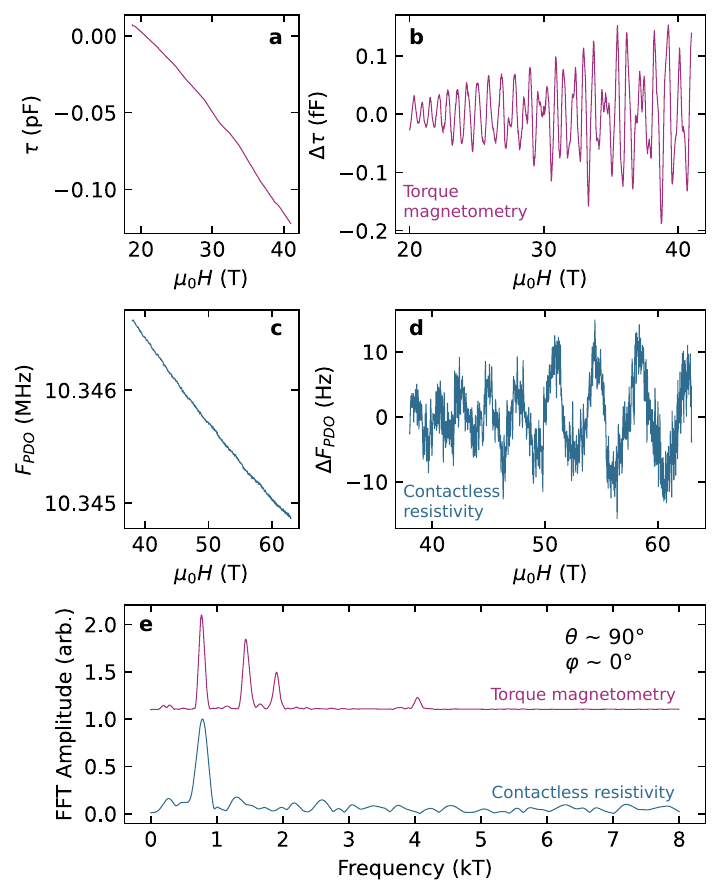} 
    \end{center}
    \vspace{-5pt}
    \caption*{\textbf{Extended Data Fig. 7 | Pulsed magnetic field QO measurements. a,} Capacitive torque magnetometry measured in steady fields and \textbf{b,} the background-subtracted oscillatory profile. \textbf{c,} Contactless resistivity measured by the PDO technique (see \textit{Methods}) in pulsed fields and \textbf{d,} the background-subtracted oscillatory profile. \textbf{e,} FFT traces for both measurements, performed with the same magnetic field orientation. The signal-to-noise ratio of the contactless resistivity is insufficient to resolve the high frequency components, however the peak at 0.8~kT is well resolved. This frequency branch appears robust, with no sign of a Lifshitz transition, at least up to 64~T. Furthermore, the background contactless resistivity is quite smooth, showing no sign of e.g. a reorientation of the magnetic moments.}
    \label{pic:pdo_v_torque}
\end{figure}

\begin{figure}[t!]
    \vspace{-2cm}
    \begin{center}
    \includegraphics[width=.88\linewidth]{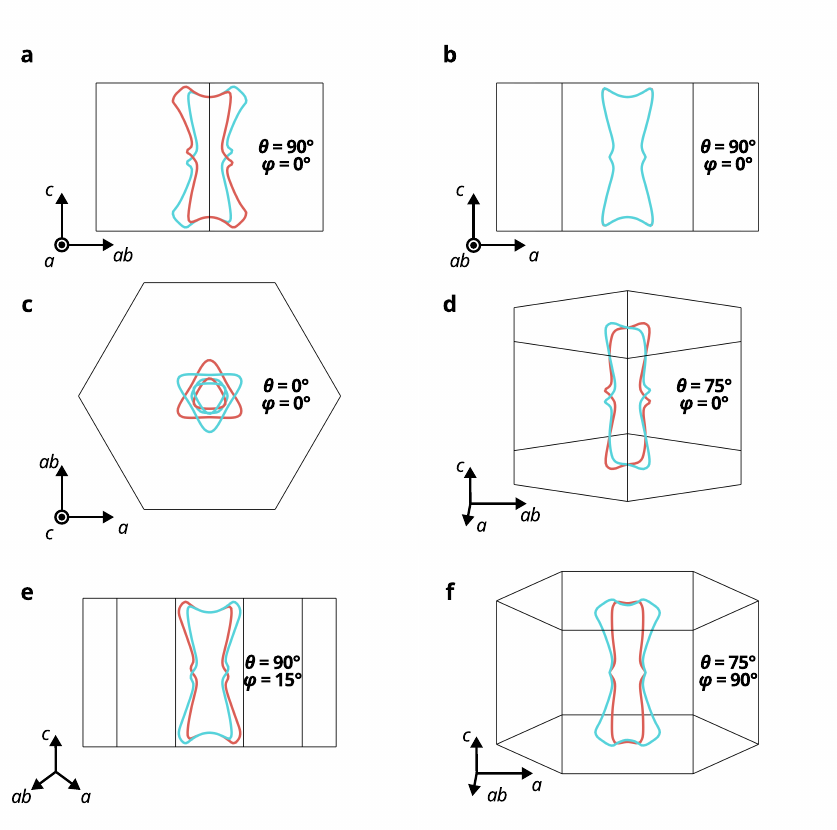} 
    \end{center}
    \vspace{-5pt}
    \caption*{\textbf{Extended Data Fig. 8 | Comparison of orbital areas for different magnetic field orientations. a,} Orienting magnetic field along the $a$-axis lies within a nodal plane, therefore the orbit for up-spins (red) can be mirrored through this plane onto a down-spin orbit. The result is that although the two orbits follow different paths in space, they are degenerate in area. Furthermore, applying fields along the \textbf{b,} $ab$-direction or \textbf{c,} $c$-direction induces quasiparticle orbits that are parallel to these nodal planes. If the up/down orbits lie in the nodal planes, they are truly degenerate in shape. If they lie above the planes, they can be mirrored to a matching orbit of the opposite spin below the plane. Magnetic fields applied within the \textbf{d,} $c-a$ and \textbf{e,} $a-ab$ planes lie along nodal directions. This produces orbits that follow different paths depending on the spin direction but are degenerate in area. \textbf{f,} Importantly, when the field is applied away from a nodal direction, but is not perpendicular to a nodal plane, we cannot use the `reflective' symmetry that spin orbits hold about nodal planes to map up-spin orbits to down-spin orbits of equal area. Instead, quasiparticles orbiting around up- and down-sheets follow different paths of different areas. This results in QOs of different frequencies, due to altermagnetic band splitting.}
    \label{pic:orbs}
\end{figure}

\begin{figure}[t!]
    \vspace{-2cm}
    \begin{center}
    \includegraphics[width=.88\linewidth]{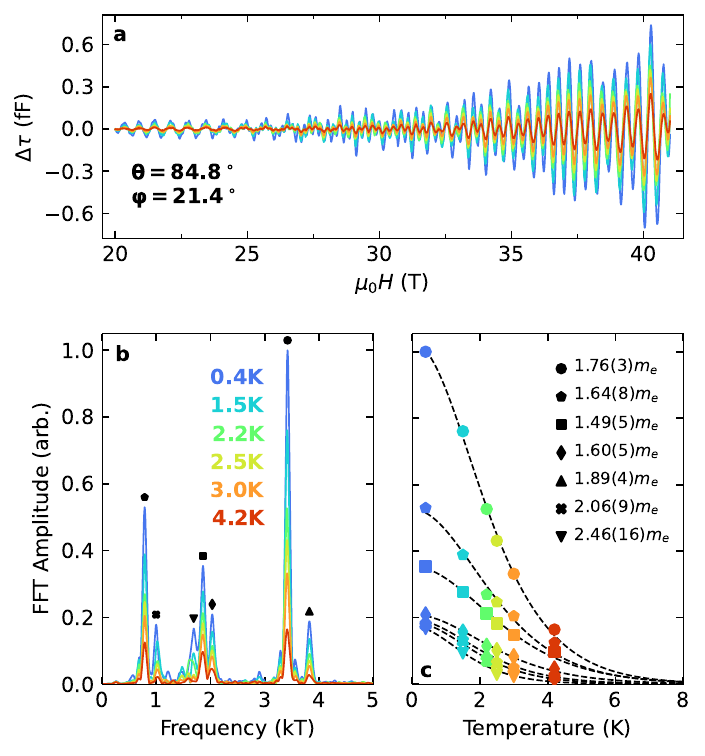} 
    \end{center}
    \vspace{-5pt}
    \caption*{\textbf{Extended Data Fig. 9 | Effective mass study.} Fits to the Lifshitz-Kosevich temperature damping of quantum oscillatory amplitude. In Fig.~\ref{Fig4} we high-pass filtered $\Delta \tau$ to focus on the Fourier components $>$~3~kT that correspond to the dogbone pocket. Here we present the non-filtered background-subtracted torque signal, which includes significant spectral weight at lower frequencies due to the web sheet. We identify seven distinct peaks in the FFT spectrum (marked with symbols), which we fitted to the LK formula~\cite{Shoenberg1984,LK}. All peaks exhibit similar cyclotron effective masses, in the range 1.5~$\leq \nicefrac{m^*}{m_e} \leq$~2.5.}
    \label{pic:extended_lk}
\end{figure}

% \newpage
% \begin{figure}[t!]
%     \vspace{-2cm}
%     \begin{center}
%     \includegraphics[width=.88\linewidth]{Extended/CrSb_dingle.pdf} 
%     \end{center}
%     \vspace{-5pt}
%     \caption*{\textbf{Extended Data Fig.  | Dingle Analysis} }
%     \label{pic:extended_Dingle}
% \end{figure}

\end{document}